\documentclass[sn-mathphys,pdflatex]{sn-jnl}

\jyear{2022}%

\theoremstyle{thmstyleone}%
\newtheorem{theorem}{Theorem}
\theoremstyle{thmstyletwo}%

\theoremstyle{thmstylethree}%
\newtheorem{definition}{Definition}%
\newtheorem {corollary} {Corollary}[section]
\usepackage[icelandic,english]{babel}
\usepackage[T1]{fontenc}

\usepackage{custom}
\usepackage{graphicx}
\graphicspath{{./}{Graphics/}}

\usepackage{booktabs}
\usepackage{array} 
\newcolumntype{L}[1]{>{\raggedright\let\newline\\\arraybackslash\hspace{0pt}}m{#1}}
\newcolumntype{C}[1]{>{\centering\let\newline\\\arraybackslash\hspace{0pt}}m{#1}}
\newcolumntype{R}[1]{>{\raggedleft\let\newline\\\arraybackslash\hspace{0pt}}m{#1}}

\usepackage{url}

\usepackage{amssymb}
\usepackage{xcolor}
\usepackage{amsmath,amsthm,amsfonts, braket, mathrsfs, mathtools, algorithm}
\usepackage{graphicx}
\usepackage{dcolumn}
\usepackage{bm}
\usepackage{physics}
\usepackage{csquotes}

\newcommand{\defeq}{\vcentcolon =}

\newcommand{\Hmin}{H_\infty}
\newcommand{\Hmax}{H_{max}}

\newcommand{\matnorm}[1]{{\left\vert\kern-0.25ex\left\vert\kern-0.25ex\left\vert #1 
    \right\vert\kern-0.25ex\right\vert\kern-0.25ex\right\vert}}

\graphicspath{ {./images/} }

\begin{document}
\title{The Trivial Bound of Entropic Uncertainty Relations}


\author*[1]{\fnm{Minu J.} \sur{Bae}}\email{minwoo.bae@uconn.edu}

\affil*[1]{\orgdiv{Department of Computer Science}, \orgname{University of Connecticut}, \orgaddress{\city{Storrs}, \postcode{06269}, \state{CT}, \country{USA}}}


\abstract{\textit{Entropic uncertainty relations} are underpinning to compute the quantitative security bound in quantum cryptographic applications, such as quantum random number generation (QRNG) and quantum key distribution (QKD). All security proofs derive a relation between the information accessible to the legitimate group and the maximum knowledge that an adversary may have gained, Eve, which exploits \textit{entropic uncertainty relations} to lower bound Eve's uncertainty about the raw key generated by one party, Alice. The \textit{standard entropic uncertainty relations} is to utilize the smooth min- and max-entropies to show these cryptographic applications' security by computing the overlap of two incompatible measurements or positive-operator valued measures (POVMs). This paper draws one case of the POVM-versioned \textit{standard entropic uncertainty relation} yielding the trivial bound since the maximum overlap in POVMs always produces the trivial value, \enquote{one}. So, it fails to tie the smooth min-entropy to show the security of the quantum cryptographic application.}


\maketitle

\section{Introduction}\label{sec:introduction}
In 1927, Heisenberg first introduced his notable uncertainty principle \cite{heisenberg1985anschaulichen}. In 1929, Robertson recomposed the principle: the product of the standard deviations of the outcomes through two incompatible orthogonal measurements on a pure quantum state is lower bounded \cite{robertson1929uncertainty}. \textit{Entropic uncertainty relations} were introduced for a finite output alphabet in \cite{hirschman1957note, deutsch1983uncertainty}. Maassen and Uffink showed  in \cite{maassen1988generalized} extended \textit{entropic uncertainty relations}, which reveals that Shannon's entropy of two incompatible measurements is lower bounded by a maximum \textit{overlap} of these measurements, which projective measurements decide the \textit{overlap} value. \textit{Entropic uncertainty relations} have developed to describe what an adversarial party with a quantum memory can entangle with a quantum system and measure the entangled system to foreknow the outcome of measurements in \cite{berta2010uncertainty, coles2011relative, renes2012one, tomamichel2011uncertainty}. Conditional von Neumann entropies craft the \textit{entropic uncertainty relations} for the presence of quantum memory. \textit{Entropic uncertainty relations} have extended to the case of general quantum measurement by a positive operator-valued measure (POVM) in \cite{krishna2002entropic}. The \textit{standard entropic uncertainty relation} with POVMs was introduced in \cite{tomamichel2011uncertainty}, namely:
\begin{equation}\label{eq:standard-ent-povms}
H_{\infty}^{\epsilon}(Z\mid  E)+H_{max}^{\epsilon}(X\mid  B)\geq -\log_{2} c,
\end{equation} 
where the overlap is defined:
\begin{equation}\label{eq:overlap}
c = \max_{a,b}\matnorm{\sqrt{M_{a}}\sqrt{N}_{b}}_{op}^{2}
\end{equation}
with $\matnorm{\cdot}_{op}$ is the \textit{operator norm} of any matrix, and $\{M_{a}\}$ and $\{N_{b}\}$ are two sets of POVMs. The inequality in Eq. (\ref{eq:standard-ent-povms}) can be interpreted as follows: if Bob, B, undoubtedly has certainty on the $X$ measurement, then Eve, E, has uncertainty to predict the outcome of the $Z$ measurement. For further information regarding \textit{entropic uncertainty relations}, we guide the reader to consider the following survey in \cite{coles2017entropic}.

In quantum cryptography, the \textit{standard entropic uncertainty relations} have been supported in a security analysis of quantum cryptographic applications, such as quantum random number generation (QRNG) and quantum key distribution (QKD). Security proof emanates from finding a relation between the information accessible to the authorized players and the full knowledge that may have been gained by an adversary, Eve. It exploits \textit{entropic uncertainty relations} to lower bound Eve's uncertainty about the raw key or random bits generated by one party, Alice. For example, the source-independent QRNG (SI-QRNG) makes use of the \textit{standard entropic uncertainty relation} with two POVMs to prove the security of the protocol in \cite{vallone2014quantum}. The source-independent means the protocol does not trust the randomness source. The brief protocol of the SI-QRNG is as follows: the adversarial source prepares the $N=m+n$ number of quantum states, $\rho_{S}$, where $\rho_{S} = tr_{E}(\rho_{SE})$. Then it sends the quantum system $\rho_{S}$ to Alice, who generates a string of true random bits. Alice randomly selects between two POVMs to measure the quantum system $\rho_{S}$, e.g, she measures the $m$-number of quantum states in the POVM measurement $\mathcal{X} = \{M_{x}\}$ and measures the $n$-number of states in the POVM measurement $\mathcal{Z} = \{N_{z}\}$. The $\mathcal{X}$ measurement is to compute a fraction of disagreement from an honest quantum state $\ket{\psi_{0}}$. And the $\mathcal{Z}$ measurement is to have a raw string of random bits. After these measurement processes, the entire quantum system will have a classical-quantum state $\rho_{ZE}$, then consider $\sigma_{YE}$ is the result of a privacy amplification process on the $Z$ register of the state. By randomly choosing a two universal hash function, the process maps the $Z$ register on the $Y$ register, namely:
\begin{equation}\label{eq:cq_state}
\sigma_{YE} = \sum_{y}\ketbra{y}_{Y} \otimes \rho_{E}^{(y)}.
\end{equation}
When the output from $Y$ is $\ell$-bit long, the real and ideal relation is shown in Eq. (\ref{eq:true-ran-bit}). Also, the length of true random bits can be approximated as follows \cite{renes2012one}:
\begin{equation}
\ell \approx H_{\infty}^{\epsilon}(Z\mid  E) - H_{max}^{\epsilon}(X).
\end{equation}
We can compute the length of the true bits $\ell$ by finding a bound of the smooth-min entropy $H_{\infty}^{\epsilon}(Z\mid  E)$. In \cite{tomamichel2011uncertainty}, the smooth-min entropy $H_{\infty}^{\epsilon}(Z\mid E)$ is bounded as follows:
\begin{equation}
H_{\infty}^{\epsilon}(Z\mid E) \geq n(-\log c) - H_{max}^{\epsilon}(X),
\end{equation}
where $c = \max_{x,z}\matnorm{\sqrt{M_{x}}\sqrt{N}_{z}}_{op}^{2}$. Also, the fraction $\mathcal{Q}$ of disagreement from an honest quantum state is estimated as follows:
\begin{equation}
H_{max}^{\epsilon}(X) \leq n\cdot h(\mathcal{Q}),
\end{equation}
where $h(\cdot)\in [0,1]$ denotes the binary entropy and $n$ is the number of random bits by a random variable $Z$. So, the length of true random bits is evaluated as follows:
\begin{equation}\label{eq:lenth-true-rand-bit}
\ell \approx n[\vartheta - 2h(\mathcal{Q})],
\end{equation}
where $\vartheta = -\log c$. Note that when the maximum measurement overlaps $c$ is always \enquote{one}, a true random bit rate, $\ell/N$, can not be positive. So, computing the overlap value of POVMs is critical for security analysis in the quantum cryptographical application. We guide the reader to consider survey paper \cite{portmann2022security} for more information about security in quantum cryptography.


Moreover, \textit{quantum algorithms} can be utilized for cryptography, search and optimization, simulation of quantum systems, and solving large systems of linear equations \cite{mosca2008quantum, montanaro2016quantum, cerezo2021variational, bharti2022noisy}. Particularly, the \textit{quantum walk} (QW), a quantum analogy of the classical random walk, is one of the fundamental methods in quantum computation and versatile quantum platforms for entanglement generation \cite{aharonov2001quantum, bednarska2003quantum, childs2003exponential, childs2009universal, lovett2010universal, li2021entangled}. Especially in the high-dimensional quantum entangled state, finding an appropriate experimental apparatus is complicated concerning the practical realization of the entangled state \cite{li2021entangled}. However, the physical implementation of QW has been discovered in many different physical systems—trapped atoms, trapped ions, and photonic systems \cite{karski2009quantum, tang2018experimental, tamura2020quantum}. Over and above that, the QW-algorithm or state can be employed not only for searching and simulating problems but also recently in quantum cryptography, e.g., QRNG, QKD, and Quantum Secure Direct Communication (QSDC) protocols \cite{rohde2012quantum, vlachou2015quantum, vlachou2018quantum, sarkar2019multi, srikara2020quantum, bae2021semi, bae2022quantum}. Particularly, quantum walk-based random number generation (QW-QRNG) protocols are introduced in \cite{sarkar2019multi, bae2021semi, bae2022quantum}. The first quantum walk-based QRNG (QW-QRNG) is introduced in \cite{sarkar2019multi}, which can generate multiple bits by only one quantum walk state. The first security analysis of the semi-source independent QW-QRNG (SI-QW-QRNG) protocols based on its original model is established in the following papers \cite{bae2021semi, bae2022quantum}. Note that the semi-source independent (SI) model assumes measurement devices are characterized, and the dimension of the system is known, but the source is under the control of an adversarial party. In the SI model, a user desires to generate true random bits without trusting the possible randomness source under the adversary's control. The SI-QW-QRNG protocol's security analysis employs the \textit{sampling-based entropic uncertainty relations} in \cite{krawec2019quantum, krawec2020new} because the \textit{standard entropic uncertainty relation} does not work with the specific sets of POVMs in the protocol; the \textit{overlap} generates the trivial value, \enquote{one}, so its \textit{entropic uncertainty relation} produces only trivial bounds in the security analysis, and the true random bit rate is always non-positive, namely:
\begin{equation}
\ell \approx n\cdot (-\log c) - 2n\cdot h(\mathcal{Q}) \leq 0,
\end{equation} 
where the maximum overlap $c$ is always \enquote{one}, which induces $-\log 1 = 0$. So, The SI-QW-QRNG protocol could not show the positive true random bit rate throughout the \textit{standard entropic uncertainty relation} with the POVMs \cite{bae2021semi}. Particularly, this paper examines and evaluates rigorously the trivial bound of the \textit{standard entropic uncertainty relation} with the POVMs, which we briefly introduce above. And it grows to gauge the newly-developed \textit{sampling-based Entropic Uncertainty relation} over more eclectic conditions to see vanquishing the trivial bound matter in \cite{bae2021semi}.

We make the following contributions to this paper. First, we introduce the analysis of the trivial value, \enquote{one}, in computing the \textit{overlap} of the POVMs in Eq. (\ref{eq:overlap-povms}) of the \textit{standard entropic uncertainty relation}. Secondly, we extend the result of the trivial value, \enquote{one}, brings about the trivial bound of the POVM-versioned \textit{standard entropic uncertainty relation} in Eq. (\ref{eq:standard_uc}). So it fails to output the positive true random bits rate in a quantum cryptographic protocol, e.g., SI-QW-QRNG. Thirdly, we expand the result of the SI-QW-QRNG in \cite{bae2021semi} to evaluate more secure key rates over various noises and dimensions of the system. 

\section{Preliminaries}
\subsection{Notation and Definitions}
This section delivers definitions and notations that will be often employed all over this paper. A density operator is a Hermitian positive semi-definite operator of unit trace functioning on a Hilbert space $\mathcal{H}$. When a pure quantum state $\ket{\varphi}\in\mathcal{H}$ is given, the \textit{density operator} of the pure state is denoted $\ketbra{\varphi}$.  A set of the d-dimensional alphabet indicates $\mathcal{A}_{d} = \{0,...,d-1\}$. Then consider a string of the $N$-number of words $q\in \mathcal{A}_{d}^{N}$ and a random subset $t \subseteq \{1,2,...,N\}$, in short, $[N]=\{1,2,...,N\}$. Then a new word $q_{t}$ illustrates the sub-string of $q$ indexed by $i\in t$. A string $q_{\bar{t}}$ insinuates the complement of $q_{t}$ in $q$. The \emph{relative Hamming weight} is $w_{a}(q) = wt_{a}(q)/\mid q\mid $, where Hamming weight the $wt_{a}(q) = \vert\{i : q_{i}\neq a\}\vert$. Note that if $a=0$, we forgo the subscription of it, namely, $w(q)$ and $wt(q) = \vert\{i : q_{i}\neq 0\}\vert$.

A notation $h_{d}(x)$ represents the $d$-ary entropy function, which is defined as: $h_{d} = x\log_{d}(d-1) - x\log_{d}(x) - (1-x)\log_{d}(1-x)$. Also, a definition $\bar{H}_{d}(x)$ is the \textit{extended $d$-ary entropy function} which is identical to $h_{d}(x)$ for all $x\in[0,1-1/d]$, is $0$ for all $x<0$, otherwise $1$ for all $x>1-1/d$. Suppose $\rho_{SE}$ be a quantum state that acts on an arbitrary Hilbert space $\mathcal{H}_{S}\otimes \mathcal{H}_{E}$. The \textit{conditional quantum min entropy} \cite{renner2008security} is defined as follows:
$
H_{\infty}(S\mid E)_{\rho} = \sup_{\sigma_{E}}\max \{\lambda \in \mathbb{R}: 2^{-\lambda} I_{S}\otimes \sigma_{E} - \rho_{SE}\geq 0\},
$
where $I_{S}$ is the identity operator on $\mathcal{H}_{S}$. Note that $\rho_{S}= \sum_{s}p_{s}\ketbra{s}$ means the $E$ system is trivial and the $S$ portion is classical. So, its min entropy is $H_{\infty}(S) = -\log \max_{s}p_{s}$. The \textit{smooth conditional min entropy} \cite{renner2008security} is defined as follows:
$
H_{\infty}^{\varepsilon}(S\mid E)_{\rho} = \sup_{\sigma\in \Gamma_{\varepsilon}(\rho)} H_{\infty}(S\mid E)_{\sigma},
$
where $\Gamma_{\varepsilon}(\rho) = \{\sigma : \norm{\sigma - \rho}\leq \varepsilon\}$ with $\norm{\cdot}$ is the trace distance of operator.

Finding the trace distance between a real classic-quantum state in Eq.(\ref{eq:cq_state}) and an ideal classic-quantum state is as follows. Suppose a classic-quantum state $\rho_{ZE}$ is given. Then consider $\sigma_{YE}$ is the result of a privacy amplification process on the $Z$ register of the state. Through a randomly chosen two-universal hash function, the process maps the $Z$ register on the $Y$ register. If the output $\ell$ bits is long, then the following relation was shown in \cite{renner2008security}:
\begin{equation}\label{eq:true-ran-bit}
\norm{\sigma_{YE} - I_{Y}/2^{\ell}\otimes \sigma_{E}} \leq 2^{-\frac{1}{2}(H_{\infty}^{\epsilon}(Z\mid E)_{\rho} - \ell) + 2\epsilon.}
\end{equation}
For example, in a QRNG protocol, $\sigma_{YE}$ is the real classic-quantum state after the post-processing. And the ideal classic-quantum state, $I_{Y}/2^{\ell}\otimes \sigma_{E}$, means the $\ell$-bits string is uniform and independent of the adversary $E$. Using Eq.(\ref{eq:true-ran-bit}), we can compute how the real state is close to the ideal state.

\subsection{The Operator Norm}
Let $A$ be any linear operator in $\mathcal{L}(\mathcal{H}, \mathcal{H}')$, where $\mathcal{H}$ and $\mathcal{H}'$ are Hilbert spaces. Then the \textit{operator norm} of the operator $A: \mathcal{H} \to \mathcal{H}'$ is the largest value by which $A$ can stretch an element in $\mathcal{H}$ via the linear transformation. The formal definition in \cite{horn2012matrix} is as follows:
\begin{definition}
Let $\mathcal{H}$ and $\mathcal{H}'$ be Hilbert spaces, and a linear operator $A\in\mathcal{L}(\mathcal{H}, \mathcal{H}')$. Then the matrix norm induced by the norm of the Hilbert space is called the \textit{operator norm}, namely:
\begin{equation}\label{eq:def-op-norm}
\matnorm{A}_{op} =\sup \Bigg\{\frac{\norm{A\ket{v}}}{\norm{\ket{v}}}: \ket{v}\in\mathcal{H} \text{ and } \norm{\ket{v}}\neq 0\Bigg\},
\end{equation}
where $\matnorm{\cdot}$ is a matrix norm and $\norm{\cdot}$ is a vector norm.
\end{definition}
Also, the \textit{operator norm}, $\matnorm{A}_{op}$, is the largest \textit{singular value} of $A$, that is, the largest \textit{eigenvalue} of $\sqrt{A^{*}A}$, where $A^{*}$ is the conjugate transpose of $A$. Note that since $A^{*}A$ is square, symmetric, and diagonalizable, the \textit{spectral theorem} implies that $(A^{*}A)^{1/2} = U\Lambda^{1/2} U^{*}$, where $U$ is a unitary \textit{eigenvectors matrix} and $\Lambda$ is an \textit{eigenvalues matrix} \cite{horn2012matrix}. Thus, we have that:
\begin{equation}\label{eq:op-max-singular}
\matnorm{A}_{op} =\sigma_{\max}(A) = \lambda_{\max}(\sqrt{A^{*}A}) = \sqrt{\lambda_{\max}(A^{*}A)},
\end{equation}
where $\lambda_{max}(\cdot)$ is the \textit{maximum eigenvalue} of a matrix.
The \textit{operator norm} satisfies the following properties \cite{horn2012matrix}:
\begin{align}
&\matnorm{A}_{op} \geq 0 \text{ and } \matnorm{A}_{op} = 0 \text{ iff } A=0\label{eq:positivity},\\
&\matnorm{\lambda A}_{op} = \mid \lambda\mid \matnorm{A}_{op} \text{ for any scalar } \lambda,\\
&\matnorm{A + B}_{op} \leq \matnorm{A}_{op} + \matnorm{B}_{op}\label{eq:sub-additivity},\\
&\matnorm{AB}_{op} \leq \matnorm{A}_{op}\matnorm{B}_{op}\label{eq:sub-multiplicativity}. 
\end{align}
Moreover, in \cite{horn2012matrix}, it says that unitary invariance in the \textit{singular value decomposition} allows: for any unitary $U$ and $V$,
\begin{equation}\label{eq:unitary-invariance}
\matnorm{UAV}_{op} = \matnorm{A}_{op}.
\end{equation}
Note that when the context is clear, we forgo the subscription in the \textit{operational norm} notation, namely $\matnorm{\cdot}_{op}  = \matnorm{\cdot}$.

\subsection{Entropic Uncertainty Relations}
After Heisenberg introduced his notable \textit{uncertainty principle} in  \cite{heisenberg1985anschaulichen}, Robertson reformulated the principle as the product of the standard deviations of the outcomes via two incompatible measurements on a pure quantum state $\ket{\varphi}$ is lower bounded \cite{robertson1929uncertainty}, namely:
\begin{equation}
\sigma_{X}\sigma_{Z} \geq \frac{1}{2} \mid \bra{\varphi} [\hat{X},\hat{Z}]\ket{\varphi}\mid ,
\end{equation}
where $\hat{X}$ and $\hat{Z}$ are observable of the two measurements, and its commutator is defined as $[\hat{X},\hat{Z}] = \hat{X}\hat{Z} - \hat{Z}\hat{X}$. After the stunning discoveries, \textit{entropic uncertainty relations} for finite output symbols were introduced by Hirschman and Deutsch, respectively, in \cite{hirschman1957note, deutsch1983uncertainty}. A matured \textit{entropic uncertainty relations} were shown by Maassen and Uffink \cite{maassen1988generalized}, which reveals that the Shannon entropy of two incompatible measurements $\mathcal{X}$ and $\mathcal{Z}$ is lower bounded by a \textit{maximum overlap} of them, called $c$, namely: for any quantum state $\rho_{A}$ before measurement,
\begin{equation}\label{eq:non_memory_uc}
H(X) + H(Z) \geq -\log_{2} c, \text{ where } c = \max_{x,z}\mid \braket{x}{z}\mid ^{2},
\end{equation}
where $\ketbra{x}\in\mathcal{X}$ and $\ketbra{z}\in\mathcal{Z}$. The value $c$ is the \textit{overlap} determined by these projective measurements. Note that the outcome of two projective measurements $\mathcal{X}$ and $\mathcal{Z}$ are stored in two classic registers $X$ and $Z$. Moreover, \textit{entropic uncertainty relations} have developed to determine what adversarial observer with a quantum memory can entangle with a quantum system and measure the entangled system to predict the outcome of measurements in \cite{berta2010uncertainty, coles2011relative, tomamichel2011uncertainty}. The \textit{entropic uncertainty relations} in the presence of quantum memory is formulated by conditional von Neumann entropies as follows: given a tripartite quantum system $\rho_{ABC}$ and orthogonal measurements $\mathcal{X}$ and $\mathcal{Z}$,
\begin{equation}
H(X\mid B)_{\rho} + H(Z\mid C)_{\rho} \geq -\log_{2} c.
\end{equation}
These \textit {entropic uncertainty relations} dispense lower bounds on the uncertainty of the outcomes of two incompatible measurements given side information \cite{tomamichel2012framework}.

\subsubsection{Standard Entropic Uncertainty Relations with POVMs}
In \cite{krishna2002entropic}, \textit{Entropic uncertainty relations} have expanded to the case of general quantum measurement, called a \textit{positive operator-valued measure} (POVM).  The \textit{overlap} of two POVMs is defined in \cite{krishna2002entropic, tomamichel2012framework} as follows:
\begin{definition}
Let $\mathcal{X} = \{X_{a}\}$ and $\mathcal{Z} = \{Z_{b}\}$ be two POVMs. The \textit{overlap} of these measurements is defined as follows:
\begin{equation}\label{eq:overlap-povms}
c(\mathcal{X}, \mathcal{Z}) \defeq \max_{a,b}\matnorm{\sqrt{X_{a}} \sqrt{Z_{b}}}_{op}^{2},
\end{equation}
where $\matnorm{\cdot}_{op}$ is the \textit{operator norm}, which is the \textit{largest singular value}. 
\end{definition}
Note that if $\mathcal{X}$ and $\mathcal{Z}$ are projective measurements, then the \textit{overlap} in Eq. (\ref{eq:overlap-povms}) reduces to another \textit{overlap} in Eq. (\ref{eq:non_memory_uc}), namely $c=\max_{x,z}\mid \braket{x}{z}\mid ^{2}$. The \textit{standard entropic uncertainty relation} with two POVMs and smooth min-and max-entropies is introduced in \cite{tomamichel2011uncertainty, tomamichel2012framework}, namely:
\begin{theorem}
Let $\rho_{ABC}$ be any tripartite quantum state, and $\mathcal{X} = \{X_{a}\}$ and $\mathcal{Z} = \{Z_{b}\}$ be two POVMs on the quantum register $A$. For $\varepsilon \geq 0$, we have that:  
\begin{equation}\label{eq:standard_uc}
\Hmin^{\varepsilon}(X\mid B)_{\rho} + \Hmax^{\varepsilon}(Z\mid C)_{\rho} \geq -\log_{2}c(\mathcal{X}, \mathcal{Z}).
\end{equation}
\end{theorem}
The operational meaning of the smooth min-and max-entropies in \cite{renes2012one, renner2008security, tomamichel2011leftover} is as follows: first, the smooth min-entropy $\Hmin^{\varepsilon}(X\mid B)_{\rho} $ estimates the maximal number of uniformly random bits that can be extracted from $X$, but it is independent of quantum side information $B$. This quantity is significant in quantum cryptography, where the job often takes to extract unassailable randomness from a quantum adversary; next, the smooth max-entropy, $\Hmax^{\varepsilon}(Z\mid C)_{\rho}$, evaluates the minimum number of extra bits of information about $Z$ required to reconstruct it from a quantum memory $C$.

\subsubsection{Sampling-based Entropic Uncertainty Relation with POVMs}
First, the \textit{sampling-based entropic uncertainty relation} with two projective measurements for a two-dimensional quantum system is introduced in \cite{krawec2019quantum}, which is accomplished by the quantum sampling technique \cite{bouman2010sampling}. The high-dimensional versioned \textit{sampling-based entropic uncertainty relation} with two orthonormal bases and one POVM for a weak measurement to estimate a noise on the source is presented in \cite{krawec2020new}. Then \textit{sampling-based entropic uncertainty relations} with two POVMs for a high-dimensional quantum cryptographic application is introduced in \cite{bae2021semi, bae2022quantum}. Especially to remedy the matter of the trivial value of the overlap in the security analysis of the SI-QW-QRNG, the POVM-version entropic uncertainty relation is first introduced in \cite{bae2021semi}. Here, the \textit{sampling-based entropic uncertainty relation} with the POVMs in \cite{bae2021semi} is represented as follows:
\begin{theorem}
\label{sampling-uncertainty-povms}
Let $\varepsilon > 0$, $0<\beta < 1/2$, and $\rho_{SE}$ an arbitrary quantum walk-based system acting on $\mathcal{H}_{S}\otimes \mathcal{H}_{E}$, where $\mathcal{H}_{S}\cong \mathcal{H}_{\mathcal{D}}^{\otimes(n+m)}$, $m<n$, and $\mathcal{H}_{\mathcal{D}}$ is the quantum walk's Hilbert space with $\mathcal{D} = 2P$. Let the POVM $\mathcal{W} = \{[\mathbf{w_{0}}], I-[\mathbf{w_{0}}]\}=\{W_{0}, W_{1}\}$ and the POVM $\mathcal{Z} = \{I_{C}\otimes [\mathbf{j}]\}_{j=0}^{P-1} = \{Z_{j}\}_{j=0}^{P-1}$, where $\ket{w_{0}}$ is a quantum walk state starting its evolution at $(c=0,p=0)$ and $P$ is the dimension of the walker's positional space. When a random subset $t$ of size $m$ of $\rho_{S}$ is measured in the POVM $\mathcal{W}$, it outputs $q\in\mathcal{A}_{2}^{m}$ and $\sigma(t,q)$ is the post-measurement state. Then the post-state is measured using POVM $\mathcal{Z}$ resulting in outcome $r\in\mathcal{A}_{P}^{n}$, where $n=N-m$.
It holds that: except with probability at most $\varepsilon^{1/3}$,
\begin{equation}
\Hmin^{\tilde{\varepsilon}}(Z\mid E)_{\rho} +n\cdot\frac{\bar{H}_{\mathcal{D}}(w(q)+\delta)}{\log_{\mathcal{D}}(2)}  \geq -\eta_{q}\log\gamma - 2\log\frac{1}{\varepsilon},
\end{equation}  
where $\tilde{\varepsilon} = 2\varepsilon+2\varepsilon^{1/3}$, 
\begin{equation}\label{eq:gamma}
\gamma  = \max_{z}\mathbb{P}_{W}\big(\ket{w_{0}}\to z\big),
\end{equation}
$z\in\{0,...,P-1\}$, and $\eta_{q} = (N-m)(1-w(q)-\delta)$, where $\delta$ is the sampling error:
\begin{equation}
\delta = \sqrt{\frac{(N+2)\ln(2/\varepsilon^{2})}{m\cdot N}}.
\end{equation}
\end{theorem}
With this \textit{sampling-based entropic uncertainty relation}, SI-QW-QRNG in \cite{bae2021semi} can resolve the trivial bound of the \textit{standard entropic uncertainty relation} with the POVMs. So, it shows the protocol's security and computes the positive true random bit rate; see Fig.3.
\subsection{Sampling and Quantum Walk-based QRNG with POVMs}
A \textit{quantum walk} is one of the fundamental algorithms in quantum computation \cite{childs2003exponential, childs2009universal, lovett2010universal}. The \textit{quantum walk} can be utilized to gain a substantial speedup over classical algorithms based on Markov chains \cite{montanaro2016quantum, magniez2007search, santha2008quantum, portugal2013quantum, balu2017probability}. In \cite{aharonov2001quantum, bednarska2003quantum}, it says that the \textit{quantum walk} propagates over the walker's Hilbert space $\mathcal{H}_{W} = \mathcal{H}_{C}\otimes \mathcal{H}_{P}$ over time $T\in\mathbb{Z}_{\geq 0}$, where $\mathcal{H}_{C}$ is the two-dimensional Hilbert coin space, $\mathcal{H}_{P}$ is the $P$-dimensional Hilbert position space, and the dimension of the walk state is $\mid \mathcal{H}_{W}\mid  = 2P$. Note that $P$ is the walker's positional dimension. The walker starts at any initial position state $\ket{c,x}$ on a cycle at time $t =0$, where $c\in\{0,1\}$ and $x\in\{0,...,P-1\}$. For each time $t\in T$ in walking propagation, the walker employs a unitary walk operator $W = S\cdot (H\otimes I_{P})$, where the shift operator $S$ on a cycle is defined in \cite{aharonov2001quantum, bednarska2003quantum} as follows:
\begin{equation}\label{eq:shift-operator}
S = \sum_{c}\sum_{x}\ketbra{c,x +(-1)^{c} (\text{mod } P)}{c,x},
\end{equation} 
$H$ is the Hadamard operator, and $I_{P}$ is an identity operator on the position space. The shift operator $S$ maps $\ket{0,x} \to \ket{0, x+1 (\text{mod }P)}$ and $\ket{1,x} \to \ket{1, x-1 (\text{mod }P)}$. Thus, the state of the quantum walk on a cycle over time t is obtained in \cite{aharonov2001quantum, bednarska2003quantum} as follows: for any $c\in\{0,1\}$ and $x\in\{0,...,P-1\}$,
\begin{equation}
\ket{w_{c,x}} = \sum_{k}\alpha_{k}^{(c,x)}\ket{0,k (\text{mod }P)} + \sum_{r}\beta_{r}^{(c,x)}\ket{0, r (\text{mod }P)} \label{eq:walk-state},
\end{equation}
where $\ket{w_{c,x}} = W^{T}\ket{c,x}$ and $k\in\{0,...,P-1\}$ and $r\in\{P,...,2P-1\}$. Note that when the context is clear, we forgo the superscription in the amplitudes, namely $\alpha_{k}^{(c,x)} = \alpha_{k}$ and $\beta_{r}^{(c,x)} = \beta_{r}$.

A quantum random number generation (QRNG) produces a string of true random numbers by using a quantum state that carries the inherent randomness throughout quantum processes \cite{vallone2014quantum, frauchiger2013true, avesani1801secure, li2019quantum, herrero2017quantum}. Recently, a quantum walk-based random number generation (QW-QRNG) is first introduced in \cite{sarkar2019multi}. The first rigorous security analysis of the semi-source-independent QW-QRNG (SI-QW-QRNG) is presented in \cite{bae2021semi}. 

\subsubsection{The Sampling-based SI-QW-QRNG Protocol}
The first sampling-based protocol of the SI-QW-QRNG with the POVMs is introduced in \cite{bae2021semi}. The protocol details are as follows: the randomness source is the state $\rho_{S} = tr_{E}(\rho_{SE})$ is possibly correlated with an adversarial system $E$. The source sends the $N=m+n$ number of quantum walk states to the QW-QRNG protocol. The protocol picks a random subset $t$ of them of size $\mid t\mid  = m$ and measures the samples in the POVM $\mathcal{W}$ to estimate noise in the source. The quantum sampling outputs the fraction of noise $q\in\mathcal{A}_{2}^{m}$ and the sampling error $\delta$, which sends to the post-processing. The QW-QRNG measures the remaining post-quantum walk states $\sigma(t,q)$ in the POVM $\mathcal{Z}$ to produce the $n$-number of raw random bits. The $n$-number of raw bits goes to the post-processing, which outputs the $\ell$-number of true random bits, so $Y$ represents the final true random string, see Fig.1.
\begin{figure}[t]
    \centering  
    \includegraphics[width=1\textwidth]{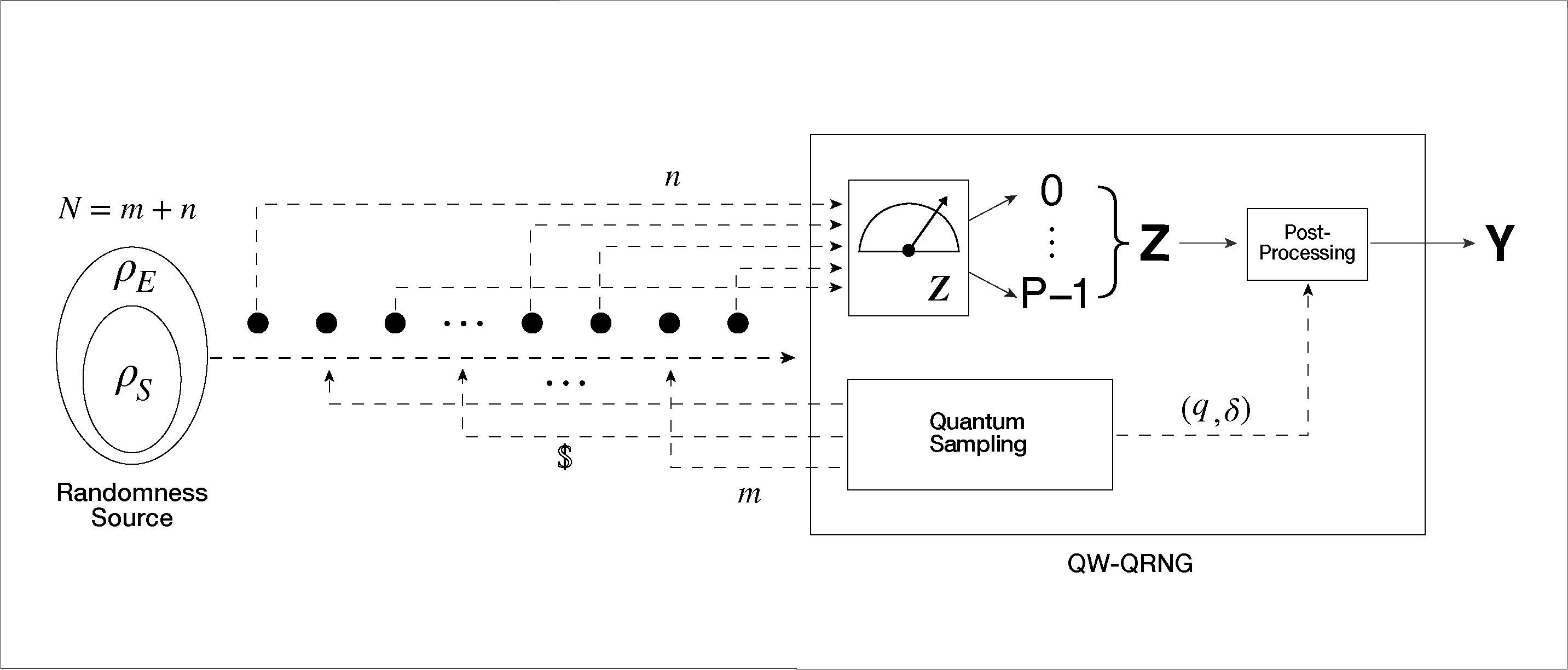}
    \caption{The sampling-based scheme of the QW-QRNG with the POVMs.}
\end{figure}
The goal of the QW-QRNG protocol is to provide that, for a given $\varepsilon_{PA}$ set by the user, after privacy amplification, the secure and random string, that is, $\varepsilon_{PA}$ close to an ideal random string, uniformly generated and independent of any adversary system, called a true random bit string.

\section{Main Result}
The SI-QW-QRNG with the POVMs, $\mathcal{W}$ and $\mathcal{Z}$, runs into the problem of estimating a true random bit rate via the \textit{standard entropic uncertainty relation} in Eq. (\ref{eq:standard_uc}) because its \textit{maximum overlap} in the POVMs always yields the trivial value, \enquote{one} \cite{bae2021semi}. Here, we formally discuss this problem. Suppose the randomness source in the SI-QW-QRNG protocol produces the honest \textit{quantum walk} state, $\ket{w_{0,0}} = W^{T}\ket{0,0}$, where $W$ is a unitary walk operator and $T$ is time. But, we assume that the randomness source may be under the adversary's control or the state's transmission is noisy. Once the \textit{quantum walk} state is created and delivered, the protocol measures the state by two POVMs, $\mathcal{W}$ and $\mathcal{Z}$, to generate a true random bit. The POVMs are defined in \cite{bae2021semi} as follows: 
\begin{equation}\label{eq:qw-povm-00}
\mathcal{W} = \{\ketbra{w_{0}}, I-\ketbra{w_{0}}\} = \{W_{0},  W_{1}\}
\end{equation}
and  
\begin{equation}\label{eq:qw-povm-01}
\mathcal{Z} = \{I_{C}\otimes\ketbra{j}\}_{j=0}^{P-1} = \{Z_{j}\}_{j=0}^{P-1}.
\end{equation}
Note that the notation $\ket{w_{0,0}}$ can be simplified as $\ket{w_{0}}$. When the protocol computes the random bit rate by finding the \textit{overlap} in Eq. (\ref{eq:overlap-povms}), it produces the trivial value, \enquote{one}. So, the \textit{standard entropic uncertainty relation} in Eq. (\ref{eq:standard_uc}) induces the trivial bound. We can go into the problem further in the following section. 

\subsection{Trivial Value of Overlaps}
First of all, we have the analytical analysis of the trivial bound of the \textit{overlap} in Eq. (\ref{eq:overlap-povms}) in the SI-QW-QRNG protocol, which shows analytically why it produces such trivial value, \enquote{one}. We formally define our main result as follows: 
\begin{theorem}
\label{main-thm}
Let $\ket{w_{0}}$ be an honest \textit{quantum walk} state on a cycle and let $\mathcal{W} = \{\ketbra{w_{0}}, I-\ketbra{w_{0}}\} = \{W_{0}, W_{1}\}$ and $\mathcal{Z} = \{I_{C}\otimes\ketbra{j}\}_{j=0}^{P-1} = \{Z_{0},..., Z_{P-1}\}$ be two POVMs, where $P\in \mathbb{Z}_{\geq 0}$ is the positional dimension of the quantum walker. Then the \textit{overlap} defined in Eq. (\ref{eq:overlap-povms}) always produces the trivial value, namely:
\begin{equation}\label{eq:trivial-value}
c(\mathcal{W}, \mathcal{Z}) = \max_{a,b}\matnorm{\sqrt{W_{a}}\sqrt{Z_{b}}}_{op}^{2} = 1,
\end{equation}
where $\matnorm{\cdot}_{op}$ is the \textit{operator norm}, that is, the \textit{largest singular value} of the given matrix.
\end{theorem}
\noindent\begin{proof}
Suppose an honest \textit{quantum walk} state is defined as $\ket{w_{0}} =\sum_{k}\alpha_{k}\ket{0,k}+\sum_{r}\beta_{r}\ket{1,r}$ and $\bra{w_{0}}=\sum_{k'}\alpha_{k'}\bra{0,k'}+\sum_{r'}\beta_{r'}\bra{1,r'}$ via Eq. (\ref{eq:walk-state}). Then we have a measurement operator $W_{0} = \ketbra{w_{0}}$ in POVM $\mathcal{W}$, namely: for any $c,c'\in\{0,1\}$,
\begin{equation}
W_{0} = \sum_{c,c'}f_{0}(c,c'),
\end{equation}
where the function $f_{0}(c,c')$ is:
\begin{equation}
f_{0}(c,c') = 
\begin{cases}
\sum_{k,k'}\alpha_{k}\alpha_{k'}^{*}\ketbra{c,k}{c',k'} & \text{if } c,c'=0,\\
\sum_{r,k'}\beta_{r}\alpha_{k'}^{*}\ketbra{c,r}{c',k'} & \text{if } c=1,c'=0,\\
\sum_{k,r'}\alpha_{k}\beta_{r'}^{*}\ketbra{c,k}{c',r'} & \text{if } c=0,c'=1,\\
\sum_{k,k'}\beta_{k}\beta_{k'}^{*}\ketbra{c,k}{c',k'} & \text{if } c,c'=1.
\end{cases}
\end{equation}

Since $W_{a}$ and $Z_{b}$ are self-adjoint operators, that is, $W_{a}^{2} = W_{a}$ and $Z_{b}^{2}= Z_{b}$, we have that $\sqrt{W_{a}}\sqrt{Z_{b}} = W_{a}\cdot Z_{b}$. Then we compute the following: $W_{0}\cdot Z_{b} = \ketbra{w_{0}}\big(I_{c}\otimes \ketbra{j}\big)$, namely: for any $b\in\{0,...,n-1\}$,
\begin{equation}
W_{0}\cdot Z_{b}  = \sum_{c,c'}f_{0}^{(b)}(c,c'),
\end{equation}
where the function is updated as follows:
\begin{equation}
f_{0}^{(b)}(c,c') = 
\begin{cases}
\sum_{k}\alpha_{k}\alpha_{b}^{*}\ketbra{c,k}{c',b} & \text{if } c,c'=0,\\
\sum_{r}\beta_{r}\alpha_{b}^{*}\ketbra{c,r}{c',b} & \text{if } c=1,c'=0,\\
\sum_{k}\alpha_{k}\beta_{b}^{*}\ketbra{c,k}{c',b} & \text{if } c=0,c'=1,\\
\sum_{r}\beta_{k}\beta_{b}^{*}\ketbra{c,k}{c',b} & \text{if } c,c'=1.
\end{cases}
\end{equation}

Let $A_{b} = W_{0}\cdot Z_{b}$. Then we compute $A_{b}^{*}A_{b}$ as follows:
\begin{equation}
A_{b}^{*}A_{b} = \sum_{c,c'}\tilde{f}_{0}^{(b)}(c,c'),
\end{equation}
where
\begin{equation}
\tilde{f}_{0}^{(b)}(c,c') = 
\begin{cases}
\alpha_{b}\alpha_{b}^{*}\ketbra{c,b}{c',b} & \text{if } c,c'=0,\\
\beta_{b}\alpha_{b}^{*}\ketbra{c,b}{c',b} & \text{if } c=1,c'=0,\\
\alpha_{b}\beta_{b}^{*}\ketbra{c,b}{c',b} & \text{if } c=0,c'=1,\\
\beta_{b}\beta_{b}^{*}\ketbra{c,b}{c',b} & \text{if } c,c'=1.
\end{cases}
\end{equation}

By the Spectral theorem, it is not difficult to find that $A_{b}^{*}A_{b}$ has only one \textit{eigenvalue}, $\lambda_{0}^{(b)}$, where $\lambda_{0}^{(b)} = \alpha_{b}\alpha_{b}^{*}+\beta_{b}\beta_{b}^{*}$. Thus, we have that:
\begin{equation}\label{eq:max-eigenval}
\max_{b}\matnorm{A_{b}}_{op}^{2}  = \max_{b} \Lambda(A_{b}^{*}A_{b}) = \max_{b} \lambda_{0}^{(b)},
\end{equation}
where $0\leq \lambda_{0}^{(b)}\leq 1$ since $\sum_{b}\alpha_{b}\alpha_{b}^{*}+\beta_{b}\beta_{b}^{*}  = 1$. Note that as the walker's positional space becomes larger, then $\lambda_{0}^{(b)}$ becomes smaller.
Now, we consider the second measurement operator in POVM $\mathcal{W}$ as follows: $W_{1} = I -W_{0}$, where $I = \sum_{i}W_{i}$ and $i=0,...,2P-1$. Then we compute the POVM $W_{1}$ as follows:
\begin{equation}
W_{1} = I - \sum_{c,c'}f_{0}(c,c')  = f_{1}(c,c').
\end{equation}

Let $B_{b} = W_{1}\cdot Z_{b}$ as follows:
\begin{equation}
f_{1}^{(b)}(c,c') = 
\begin{cases}
\sum_{k}(1-\alpha_{k}\alpha_{b}^{*})\ketbra{c,k}{c',b} & \text{if } c,c'=0,\\
\sum_{r}-\beta_{r}\alpha_{b}^{*}\ketbra{c,r}{c',b} & \text{if } c=1,c'=0,\\
\sum_{k}-\alpha_{k}\beta_{b}^{*}\ketbra{c,k}{c',b} & \text{if } c=0,c'=1,\\
\sum_{r}(1-\beta_{r}\beta_{b}^{*})\ketbra{c,k}{c',b} & \text{if } c,c'=1.
\end{cases}
\end{equation}
Then we compute $B_{b}^{*}B_{b}$ as follows:
\begin{equation}
B_{b}^{*}B_{b} = \sum_{c,c'}f_{1}^{(b)}(c,c'),
\end{equation}
where $c,c'\in\{0,1\}$ and 
\begin{equation}
\tilde{f}_{1}^{(b)}(c,c') = 
\begin{cases}
(1-\alpha_{b}\alpha_{b}^{*})\ketbra{c,b}{c',b} & \text{if } c,c'=0,\\
-\beta_{b}\alpha_{b}^{*}\ketbra{c,b}{c',b} & \text{if } c=1,c'=0,\\
-\alpha_{b}\beta_{b}^{*}\ketbra{c,b}{c',b} & \text{if } c=0,c'=1,\\
(1-\beta_{b}\beta_{b}^{*})\ketbra{c,b}{c',b} & \text{if } c,c'=1.
\end{cases}
\end{equation}
Since $B_{b}^{*}B_{b}$ is square and symmetric, we can have the decomposition as follows:
\begin{equation}
B_{b}^{*}B_{b} = Q_{b}\Lambda(B_{b}^{*}B_{b}) Q_{b}^{*},
\end{equation}
where 
\begin{equation}
Q_{b} = \frac{1}{\sqrt{\lambda_{0}^{(b)}}}\sum_{c,c'}g_{1}^{(b)}(c,c'),
\end{equation}
where $\lambda_{0}^{(b)}$ is as same as the \textit{maximum eigenvalue} of $A_{b}^{*}A_{b}$ in Eq. (\ref{eq:max-eigenval}), 
\begin{equation}
g_{1}^{(b)}(c,c') = 
\begin{cases}
-\beta_{b}\ketbra{c,b}{c',b}        & \text{if } c, c'=0,\\
\alpha_{b}^{*}\ketbra{c,b}{c',b}   & \text{if } c=1,c'=0,\\
\alpha_{b}\ketbra{c,b}{c',b}         & \text{if } c=0,c'=1,\\
-\beta_{b}^{*}\ketbra{c,b}{c',b}  & \text{if } c,c'=1,\\
\end{cases}
\end{equation}
and the \textit{eigenvalue matrix} is formed as follows:
\begin{equation}
\Lambda(B_{b}^{*}B_{b}) = 1\cdot\ketbra{0,b}{0,b}+ (1-\lambda_{0}^{(b)})\ketbra{1,b}{1,b}.
\end{equation}
So, we have the following result: an \textit{eigenvalue} of the symmetric matrix, $B_{b}^{*}B_{b}$, is $\lambda_{1}^{(b)}\in \{1, 1-\lambda_{0}^{(b)}\}$. Since the \textit{eigenvalue} $\lambda_{0}^{(b)}$ is always less than equal to one, we have that:
\begin{equation}
\max_{b}\matnorm{W_{1}\cdot Z_{b}}_{op}^{2}  = \max_{b} \lambda_{1}^{(b)}= 1.
\end{equation}
Therefore, the maximum of the \textit{operator norm} overall measurements in two POVMs $\mathcal{W}$ and $\mathcal{Z}$ produces always the trivial value, one, as claimed.
\end{proof}

\begin{corollary}\label{trivial-standard-uc}
Let $\rho_{SE}$ be an arbitrary \textit{quantum walk} state acting on $\mathcal{H}_{S}\otimes\mathcal{H}_{E}$. And let $\mathcal{W} = \{W_{0}, W_{1}\}$ and $\mathcal{Z} = \{Z_{j}\}_{j=0}^{P-1}$ be two POVMs on the quantum register $S$. Then the \textit{standard entropic uncertainty relation} in Eq. (\ref{eq:standard_uc}) induces the trivial bound. 
\end{corollary}
We have this following consequence because its \textit{overlap} in Eq. (\ref{eq:overlap-povms}) shows the trivial value by Theorem \ref{main-thm}. Namely, the \textit{standard entropic uncertainty relation} in Eq. (\ref{eq:standard_uc}) with the POVMs $\mathcal{W}$ in (\ref{eq:qw-povm-00}) and $\mathcal{Z}$ in (\ref{eq:qw-povm-01}) shows:  
\begin{equation}
H_{\infty}^{\epsilon}(Z\mid E)+H_{max}^{\epsilon}(W)\geq -\log_{2} 1 = 0,
\end{equation}
which is lower-bounded by the trivial hurdle, ``zero." Moreover, we can estimate the length of the true random bits via Eq. (\ref{eq:lenth-true-rand-bit}) as follows:
\begin{equation}
\ell \approx n\cdot 0 - 2n\cdot h(\mathcal{Q}) \leq 0,
\end{equation}
which always outputs the non-positive length of true random bits. Thus, the \textit{sampling-based entropic uncertainty relation} with the POVMs is introduced in \cite{bae2021semi} to show the security of the SI-QW-QRNG.

\section{Evaluations}
From Theorem \ref{main-thm}, we show that the \textit{overlap} $c(\mathcal{W}, \mathcal{Z})$ in Eq. (\ref{eq:overlap-povms}) with the POVMs $\mathcal{W}$ and $\mathcal{Z}$ in Eq. (\ref{eq:qw-povm-00}) and Eq. (\ref{eq:qw-povm-01}) always outputs the trivial value, \enquote{one}. We newly simulate this result in this section to show the trivial bound consequence is valid over all dimensions and times. To have the experimental results, we define functions as follows: 
\begin{equation}\label{eq:delta-00}
\delta_{0}(\mathcal{W}, \mathcal{Z}) = \max_{b}\matnorm{\sqrt{W_{0}}\sqrt{Z_{b}}}_{op}^{2}
\end{equation}
and 
\begin{equation}\label{eq:delta-01}
\delta_{1}(\mathcal{W}, \mathcal{Z})  = \max_{b}\matnorm{\sqrt{W_{1}}\sqrt{Z_{b}}}_{op}^{2},
\end{equation}
which computes the \textit{overlap} $c(\mathcal{W}, \mathcal{Z}) = \max\{\delta_{0}, \delta_{1}\}$, see Fig.2. 

Moreover, we evaluate more key rates with the new \textit{sampling-based relation} over different noises, e.g., $Q=0$, $0.15$, and $0.20$, and varied dimensions of the system, e.g., $\mathcal{D} = 2\cdot 3$, $2\cdot 5$, $2\cdot 11$,  $2\cdot 21$, $2\cdot51$, than the previous simulation result in \cite{bae2021semi}. Note that $\mathcal{D} = 2\cdot P$, where $P$ are the walker's positional dimension.

\begin{figure}[h]\label{fig:position-simulation}
    \centering  
    \includegraphics[width=1\textwidth]{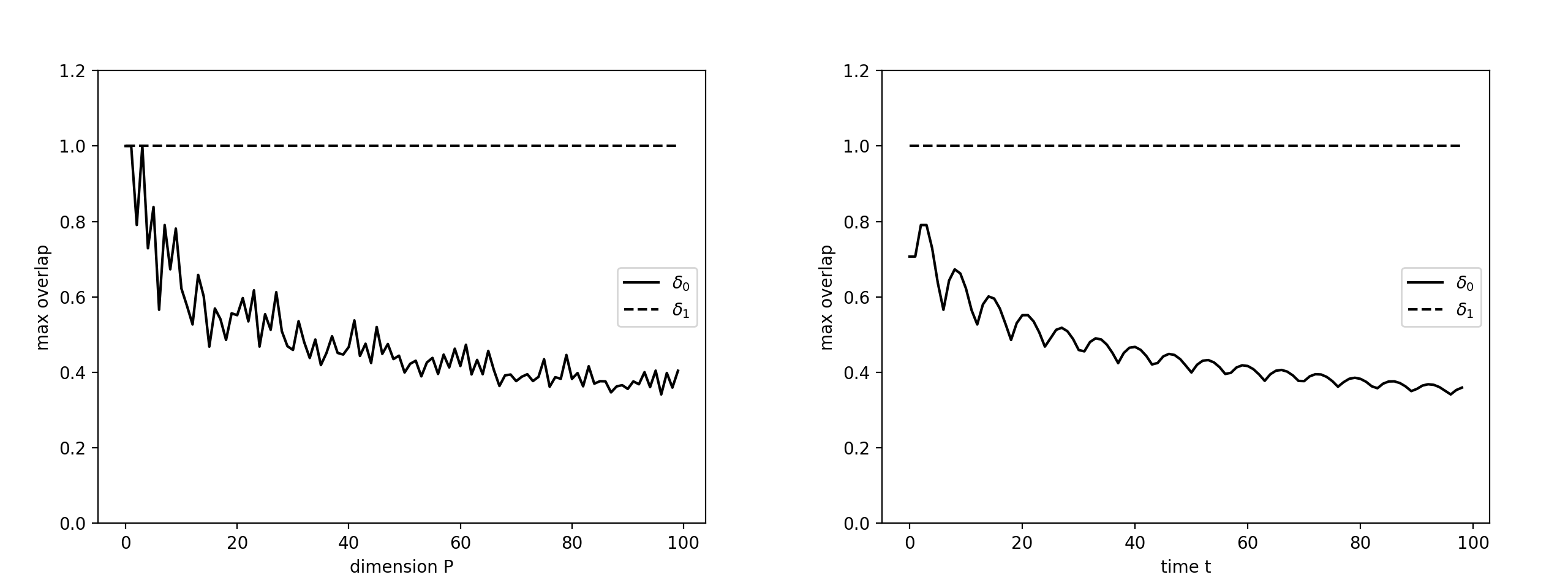}
    \caption{The left graph is the case of \textit{quantum walk} state evolves over dimensions $P = 1,...,101$ and time $T = P$; it show $\delta_0$ and $\delta_1$ to find the maximum \textit{overlap} in the \textit{standard entropic uncertainty relation}. The right graph is the case of \textit{quantum walk} state evolves over time $T=1,...,100$ for fixed dimension $P=101$; it also shows $\delta_0$ and $\delta_1$ to find the maximum \textit{overlap} in the \textit{standard entropic uncertainty relation}.}
\end{figure}


\subsection{Computing the True Random Bit Rate of the SI-QW-QRNG}
In the SI-QW-QRNG \cite{bae2021semi}, the \textit{sampling-based entropic uncertainty relation} with two POVMs $\mathcal{W}$ and $\mathcal{Z}$ is introduced to remedy the trivial bound issue the above presented. This section briefly shows the true random bit rate result using the sampling-based tool for readers interested in the issue's end result. First of all, the dimension of the quantum walk state is defined as follows $\mathcal{D} = 2P$, where $P$ are the positional dimension of the walker's state. The protocol using the \textit{sampling-based entropic uncertainty relation} with the POVMs in Theorem \ref{sampling-uncertainty-povms} can find the length of true random bits as follows: except with probability at most $\varepsilon^{1/3}$, the final secret string of size is:
\begin{equation}\label{eq:new-length-true-rand-bits}
\ell_{new} = -\eta_{q}\log_{2}\gamma - n\cdot \frac{\bar{H}_{2P}(w(q)+\delta)}{\log_{2P}(2)} - 2\log_{2}\frac{1}{\varepsilon},
\end{equation}
which is $(5\varepsilon +4 \varepsilon^{1/3})$-close to an ideal random string, i.e., one that is uniformly generated and independent of any adversarial system. The true random bit rates of the SI-QW-QRNG shows in Fig.3. For the technical details and proof of the \textit{sampling-based entropic uncertainty relations} with the POVMs, we advise the reader to consider the paper \cite{bae2021semi}.
\begin{figure}[t]
    \centering  
    \includegraphics[width=1\textwidth]{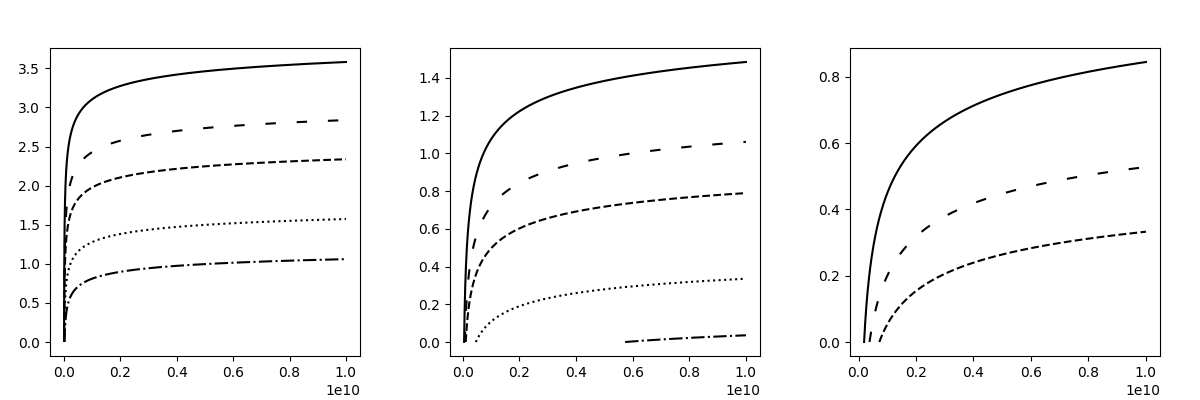}
    \caption{By using the \textit{sampling-based uncertainty relation} with the POVMs, a quantum walk-based QRNG produces true random bits over noisy settings; The left graph shows the random bit rate over noise $Q=0$; The middle graph shows the random bit rate over noise $Q=0.15$; The last graph shows the random bit rate over noise $Q=0.2$; x-axis: the $N$-number of signals sent from the source; y-axis: random bit rates, $\ell_{new}/N$, where $\ell_{new}$ is the number of true random bits via Eq.(\ref{eq:new-length-true-rand-bits}); the solid line represents $\mathcal{D} = 2\cdot 51$; the loosely dashed line is $\mathcal{D}= 2\cdot 21$; the dashed line is $\mathcal{D} = 2\cdot 11$; the dotted line is $\mathcal{D}= 2\cdot 5$; the dot-dashed line is $\mathcal{D} = 2\cdot 3$.}
\end{figure}

\section{Closing Remarks}
In this paper, we sketch the case that the \textit{standard entropic uncertainty relation} with the POVMs fails to use in security proof in the quantum cryptographic application, e.g., the SI-QW-QRNG. In this case, the \textit{standard entropic uncertainty relation} with the POVMs produces the trivial bound since its \textit{overlap} in Eq. (\ref{eq:overlap-povms}) always outputs the value, \enquote{one}. We show the analytical proof of always producing the trivial value of the \textit{overlap} in the sets of POVMs. Also, we show the simulation results of their overlaps to show it always produces the trivial value \enquote{one}. There are exciting open problems: first of all, the \textit{generalized entropic uncertainty relation} with the effective overlap $c^{*}$ for generalized measurement operators, POVMs, was introduced in \cite{coles2017entropic, tomamichel2012framework}. It is paramount to check whether the effective overlap $c^{*}$ outputs the trivial value, \enquote{one}, with the POVMs setup in Eq.(\ref{eq:qw-povm-00}) and Eq.(\ref{eq:qw-povm-01}). Secondly, we can discuss finding which sets of POVMs do not operate with the \textit{standard entropic uncertainty relations} for the security analysis of a quantum cryptographic application. Thirdly, we can investigate how to rectify the issue from the trivial bound of the standard entropic uncertainty relation in a quantum cryptographic application and secure quantum communication. Moreover, we can invent a possible alternative and new entropic uncertainty relation, e.g., the \textit{sampling-based entropy uncertainty relation}, to handle the trivial matter in the security analysis.
\section{acknowledgments}
The author would like to thank Walter O. Krawec for valuable feedback and helpful discussions to ripe the paper's quality.

\bibliography{sn-article}

\end{document}